\documentclass{article}

\PassOptionsToPackage{numbers}{natbib}

\usepackage[final]{neurips_2023}
\usepackage{subfigure}
\usepackage{graphicx}
\usepackage{amsmath}
\usepackage[symbol]{footmisc}

\usepackage[utf8]{inputenc} 
\usepackage[T1]{fontenc}    
\usepackage{hyperref}       
\usepackage{url}            
\usepackage{booktabs}       
\usepackage{amsfonts}       
\usepackage{nicefrac}       
\usepackage{microtype}      
\usepackage{xcolor}         

\title{ScripTONES: Sentiment-Conditioned Music Generation for Movie Scripts}

\author{%
  Vishruth Veerendranath, Vibha Masti, Utkarsh Gupta, Hrishit Chaudhuri, Gowri Srinivasa \\
  PES Center for Pattern Recognition, Department of Computer Science and Engineering \\
  PES University \\
  \texttt{\{vishruthnath, vsmasti, utkarsh348, hrishitchaudhuri\}@gmail.com}, \\ \texttt{gsrinivasa@pes.edu} 
}

\begin{document}

\maketitle

\begin{abstract}
  Film scores are considered an essential part of the film cinematic experience, but the process of film score generation is often expensive and infeasible for small-scale creators. Automating the process of film score composition would provide useful starting points for music in small projects. In this paper, we propose a two-stage pipeline for generating music from a movie script. The first phase is the Sentiment Analysis phase where the sentiment of a scene from the film script is encoded into the valence-arousal continuous space. The second phase is the Conditional Music Generation phase which takes as input the valence-arousal vector and conditionally generates piano MIDI music to match the sentiment. We study the efficacy of various music generation architectures by performing a qualitative user survey and propose methods to improve sentiment-conditioning in VAE architectures. \footnotetext{\copyright \space Veerendranath et.al \& ACM, 2024. 
 This is the author’s version of the work. It is posted here for your personal use. Not for redistribution. The Version of Record is to be published in "The Third International Conference on Artificial Intelligence and Machine Learning Systems Proceedings", \url{https://doi.org/10.1145/3639856.3639891}}
\end{abstract}

\section{Introduction}
Film scores are a critical part of the cinematic experience and in high-production environments, they require a significant amount of skill and time to produce. The challenge lies in being able to correctly identify abstract musical themes and motifs. The current state of musical score composition relies on human knowledge of these properties, with the pipeline often starting with the composers seeing a rough cut of the script, analyzing the mood of the scene, and applying this knowledge of motifs to compose music that matches the sentiment of the scene. The music is then ornamented with instrumentalization, and the score is edited to match the film shots. 

In this paper, we present \textbf{ScripTONES} (\textbf{S}cript \textbf{TO} \textbf{N}otes with \textbf{E}motional \textbf{S}ignals), an automated two-stage pipeline to conditionally generate music for movie scenes. Unlike \cite{shriramsonus}, the goal of ScripTONES is to generate music for a short movie scene, and not to retrieve music to fit the sentiment of a book chapter. The first stage of the pipeline is the \emph{Sentiment Analysis} phase where the sentiment of the movie scene is extracted from the script text. After preprocessing the scene text, we capture its sentiment by extracting its Valence and Arousal \cite{posner2005circumplex} values according to the NRC VAD lexicon \cite{vad-acl2018}. 
The second stage is conditional music generation. Here, we generate polyphonic piano MIDI music to match the valence and arousal of the scene extracted in the first stage. We explore the current transformer-based \cite{hung2021emopia, huang2018musictransformer} and VAE-based \cite{roberts2018hierarchical, wang2020pianotree, jiang2020transformervae}  methodologies that allow for sentiment-conditioned music generation. To allow for finer control from the creator in inspiring the music generated, we propose regularisation losses for sentiment modification. We also compute attribute vectors for each quadrant of the Valence-Arousal space (henceforth VA space) and perform attribute vector arithmetic to modify the sentiment of a music piece. The VAE would also allow for interpolation to match the changes in the sentiment of a scene. A detailed discussion of related work and data used can be found in Appendices \ref{app:related_work} and \ref{app:data} respectively.

\begin{figure*}
    \centering
    \includegraphics[width=0.9\textwidth]{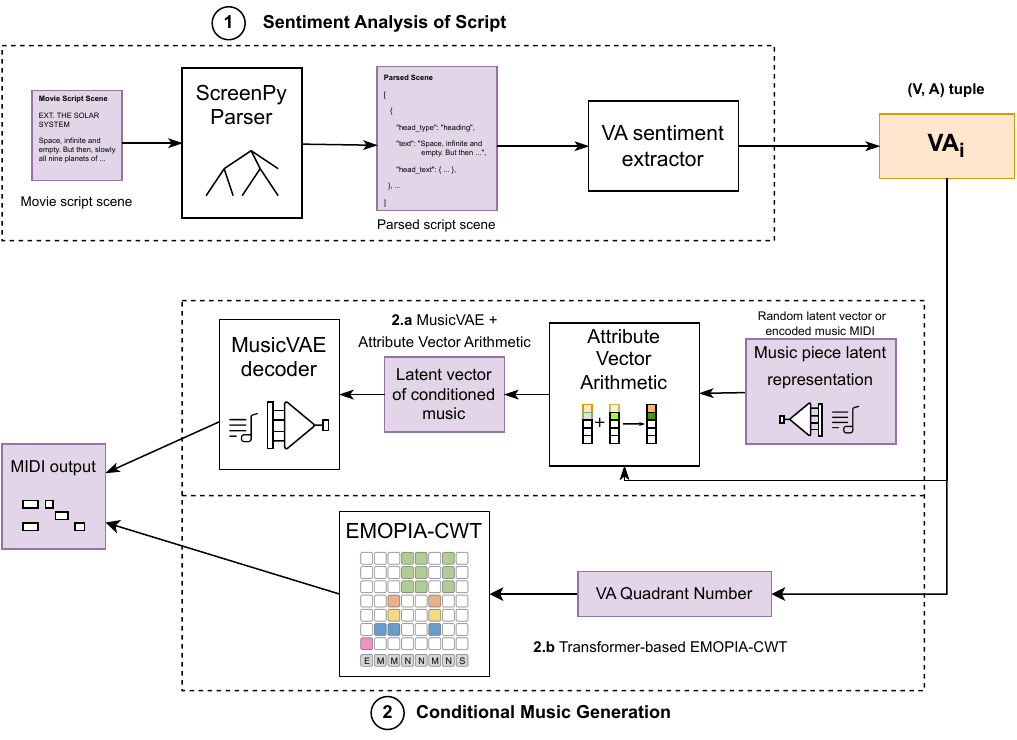}
    \caption{Illustration of the ScripTONES pipeline, which consists of 2 major stages - Sentiment Analysis of Scripts and Conditional Music Generation. Music Generation is achieved either with MusicVAE with attribute vector arithmetic or EMOPIA-CWT}
    \label{fig:pipeline}
\end{figure*}

\section{ScripTONES Pipeline}
This section describes the pipeline of score generation process which has been illustrated in Fig. \ref{fig:pipeline}. 

\subsection{Sentiment Analysis of Scripts}

A movie script contains dialogue as well as additional information that is used to produce and direct the movie. We parse the script to only extract meaningful words that contribute to the sentiment of the scene, as well as segment a script into individual scenes. For this, we use the ScreenPy parser \cite{winer2017screenpy} which relies on formal grammar rules that conform to the \emph{Hollywood Standard format}. 

For sentiment analysis, we use the NRC VAD \cite{vad-acl2018} lexicon to find the \emph{unweighted average} of valence and arousal values from every single word in the parsed movie scene. The value obtained is in the range \verb|[0, 1]|, and further normalized to be in the range \verb|[-1, 1]|. The movie scene $S_i$ is thus associated with a Valence-Arousal ($\mathit{VA}$) tuple value $\mathit{VA}_i = (V, A)$, where $-1 \le V, A \le 1$.

\subsection{Conditional Music Generation}
Below we describe the two models used in the pipeline. We propose and experiment with ways (regularization losses) to improve conditional generation with VAEs in Appendix \ref{app:improving} and \ref{app:finetune}.

\subsubsection{\textbf{EMOPIA-CWT (Transformer-based)}}
We experimented with the Compound Word Transformer (CWT) trained on EMOPIA \cite{hung2021emopia} for music generation (henceforth referred to as \textbf{EMOPIA-CWT}). This phase is labelled 2.b in Fig. \ref{fig:pipeline}. Here, sentiment is not represented as a continuous ($\mathit{V, A}$) tuple, but as a quadrant number $Q$ of the valence-arousal space in the range $[1,4]$. The EMOPIA-CWT generates music that fits the sentiment of the corresponding quadrant.

While the discrete quadrant method does not allow for latent space sampling, the music generated is quite realistic, polyphonic and fits the sentiment of movie scenes rather well.

\subsubsection{\textbf{MusicVAE tuned with Attribute Vector Arithmetic}} \label{sec:attributevec}
Inspired by \cite{roberts2018hierarchical}, we extend the idea of attribute vector arithmetic in VAEs to the sentiment attributes of valence and arousal in music. To formalize this, we define 4 emotional attribute vectors, namely High Valence ($z_{vh}$), Low Valence ($z_{vl}$), High Arousal ($z_{ah}$) and Low Arousal ($z_{al}$) by averaging latent vectors $z$ (encoded using MusicVAE \cite{roberts2018hierarchical}) of all MIDI samples in the EMOPIA dataset \cite{hung2021emopia}. To conditionally generate the music, we scale the attribute vectors based on threshold value $\alpha$ defined on the $V$ \& $A$ values extracted from the script (empirically $\alpha$ = -0.25), as per Eqn. \ref{eq:1}. This results in an Emotionally Conditioned latent vector $z_{ec}$, which is then decoded back to music as a MIDI file. This methodology enables conditioning music in a continuous space. 

\begin{equation}
z_{ec} = \begin{cases}
            |V| * z_{vh} + |A| * z_{ah} & (V \geq 0, A \geq \alpha)\\
            |V| * z_{vh} + |A| * z_{al} & (V \geq 0, A < \alpha)\\
            |V| * z_{vl} + |A| * z_{ah}  & (V < 0, A \geq \alpha)\\
            |V| * z_{vl} + |A|* z_{al}  & (V < 0, A < \alpha)\\
            \end{cases}
\label{eq:1}
\end{equation}

We introduce stochasticity by randomly sampling a point in the latent space ($z_r$) and tuning the random point based on our emotional condition vector as per $z_{tuned} = z_r + z_{ec}$. We further enable finer control for artists by replacing $z_r$ with the encoding for artist's inspiration music piece ($z_{insp}$).

\section{Evaluation and Results}
Additional experiments are detailed in Appendix \ref{sec:add_exp}. Demonstrations \& samples are on our webpage \footnote{\href{https://bit.ly/scriptones-script-music}{bit.ly/scriptones-script-music}}.

\subsection{Attribute Vector Arithmetic}
\begin{figure}[htbp]
\subfigure[Original music\label{fig:attr_og}]{\includegraphics[width=0.33\columnwidth]{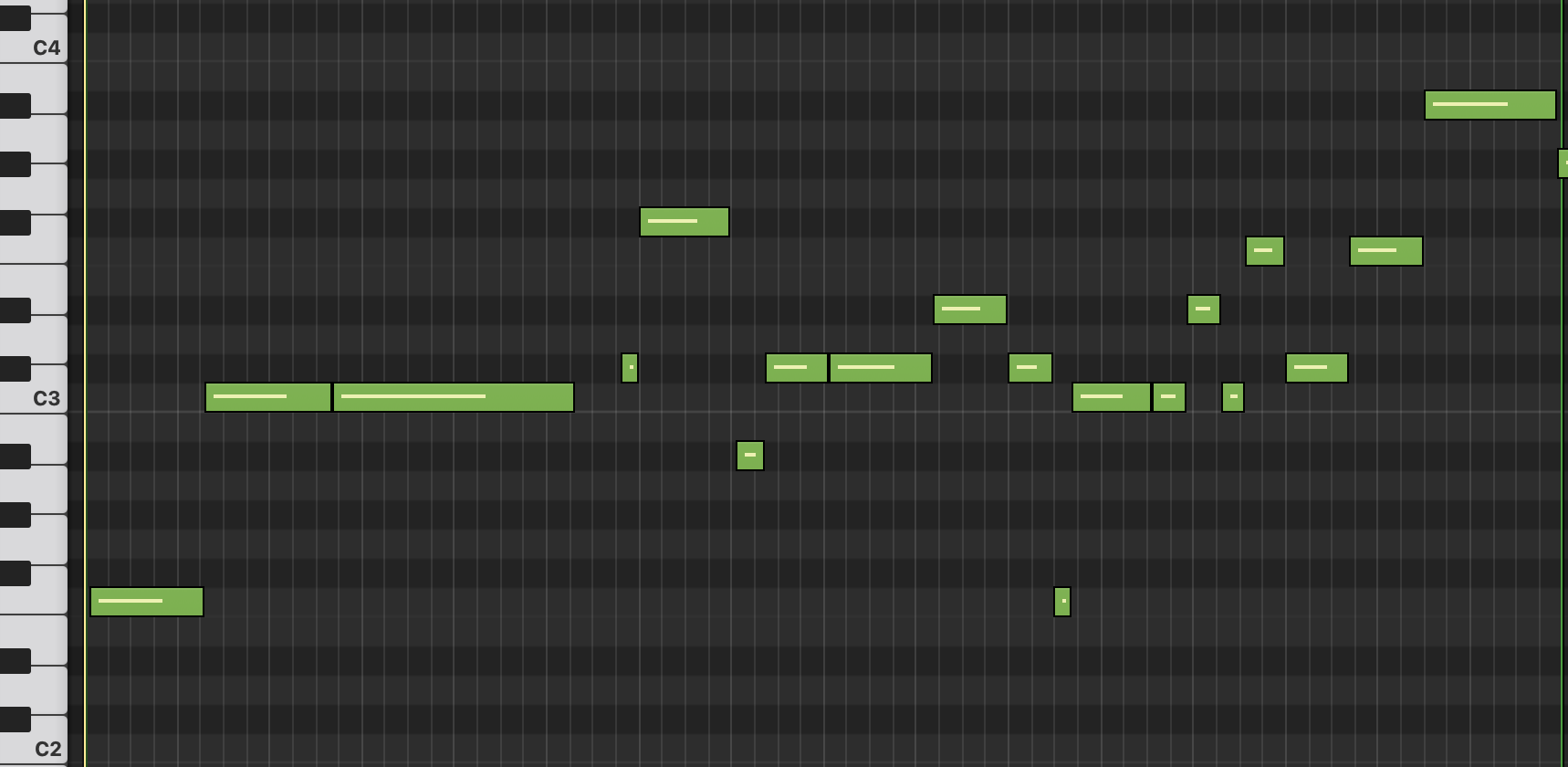}} 
\subfigure[Modified with increased valence\label{fig:attr_val}]{\includegraphics[width=0.33\columnwidth]{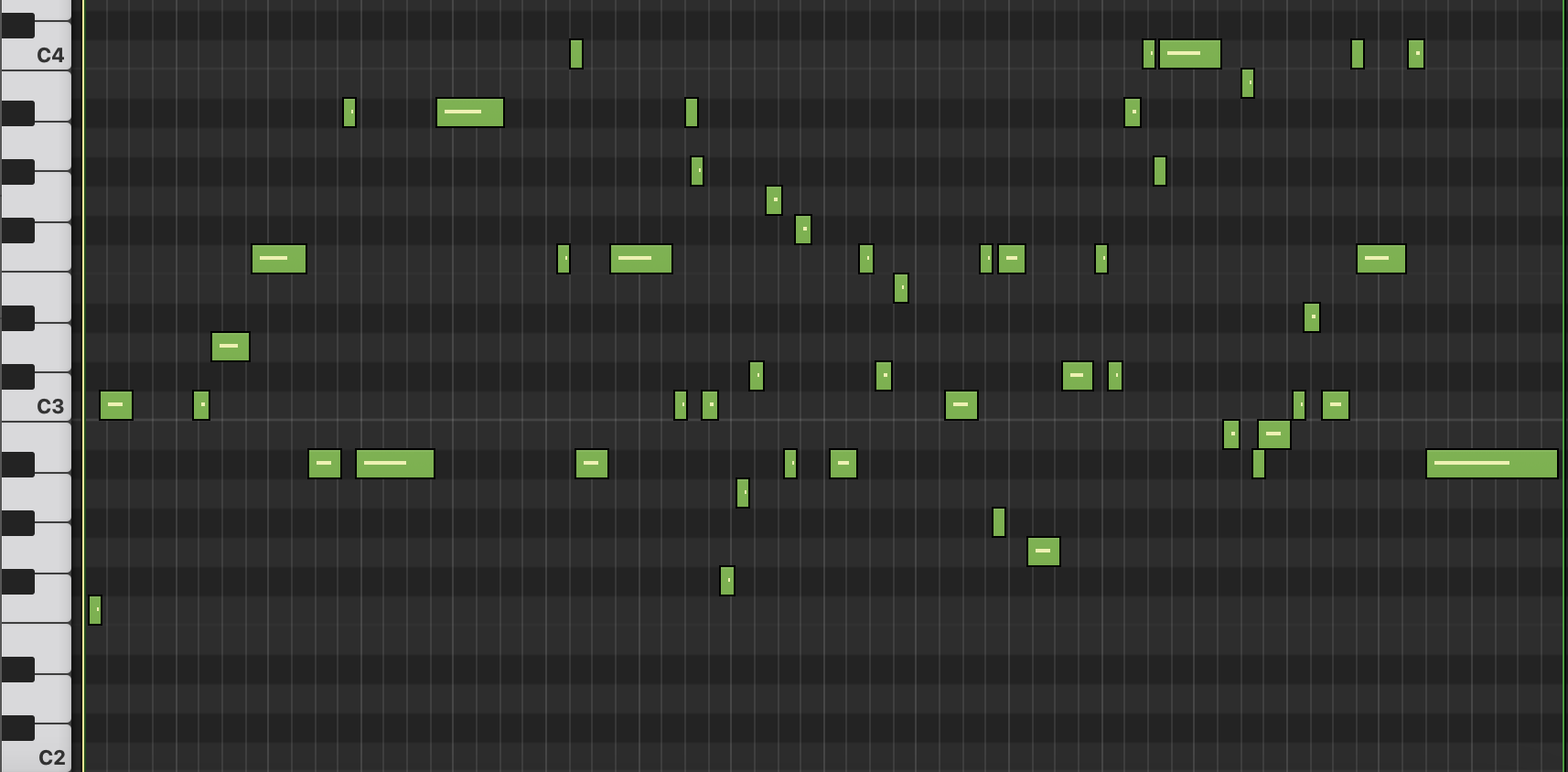}}
\subfigure[Modified with increased arousal\label{fig:attr_arou}]{\includegraphics[width=0.33\columnwidth]{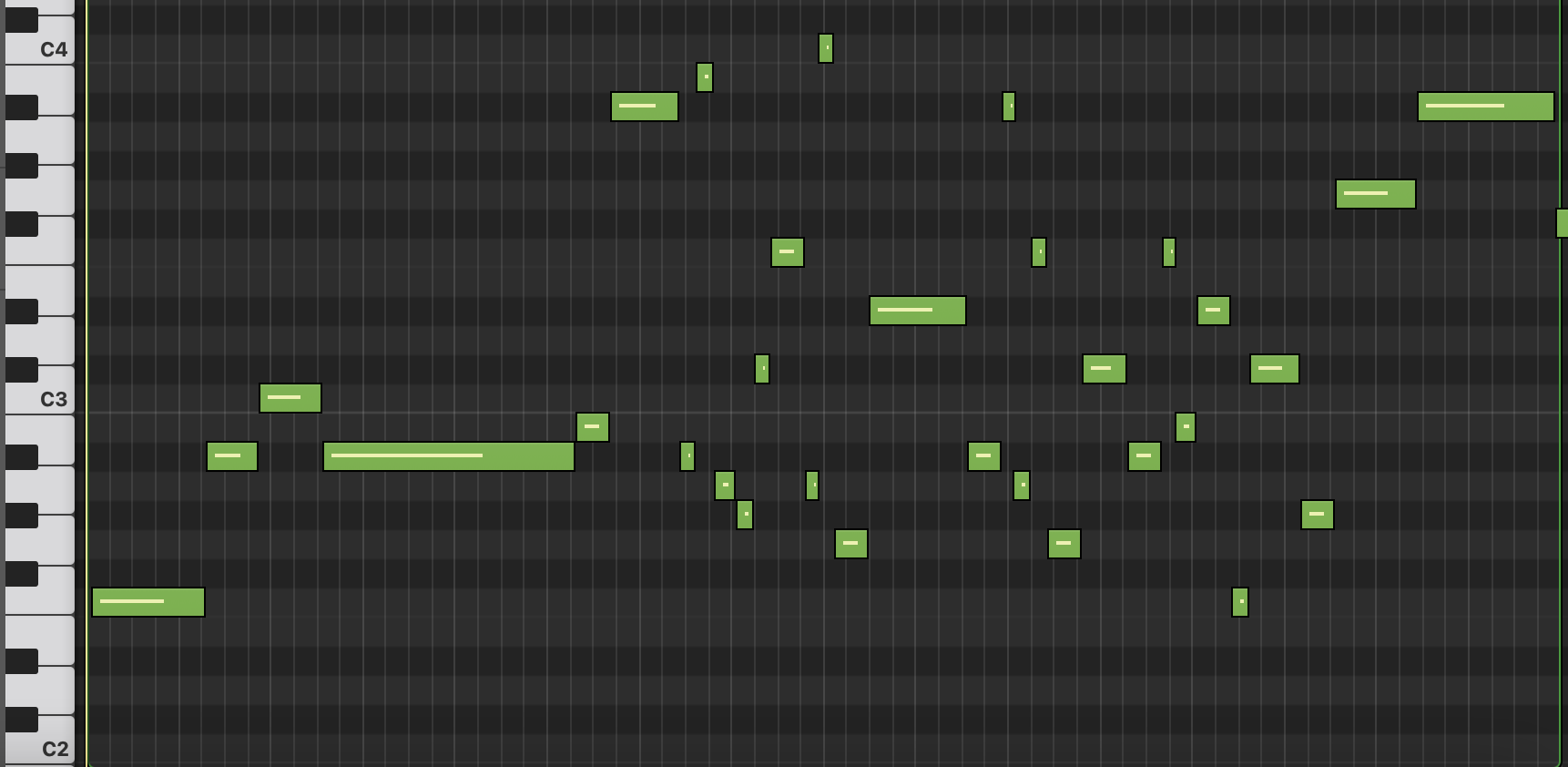}}
\caption{Original music (a) modified with Increased Valence (b) and Increased Arousal (c)}
\end{figure}

To demonstrate the validity of the attribute vector arithmetic, we show an example of conditional music generation with the pre-trained MusicVAE \cite{roberts2018hierarchical} model with attrbite vector arithmetic. A sample from the EMOPIA dataset \cite{hung2021emopia} (Fig \ref{fig:attr_og}) has been passed in as an inspiration piece and is encoded to latent vector $z$. The emotional characteristics of this piece have then been modified using the High Valence ($z_{vh}$) and High Arousal ($z_{ah}$) attribute vectors. When $z_{vh}$ is added to $z$ and then converted back to MIDI, the density of notes, as seen in Fig \ref{fig:attr_val}, is much higher than the original. Similarly, adding $z_{ah}$ to $z$ and converting back to MIDI (Fig \ref{fig:attr_arou}) causes the music to follow a similar note sequence as the original inspiration music, but intersperses it with some fast-paced staccato notes between the original note structure.

\begin{figure}[htbp]
\centering
\subfigure[Discrete Latent Regularization\label{fig:discrete_reg}]{\includegraphics[width=0.4\columnwidth]{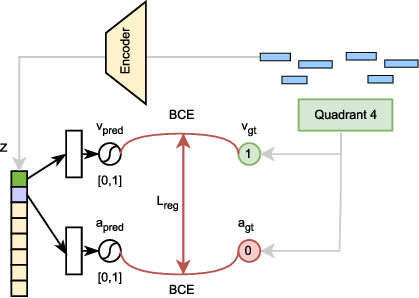}}
\hfill
\subfigure[Regularization loss during finetuning\label{fig:disc_reg_plot}]{\includegraphics[width=0.4\columnwidth]{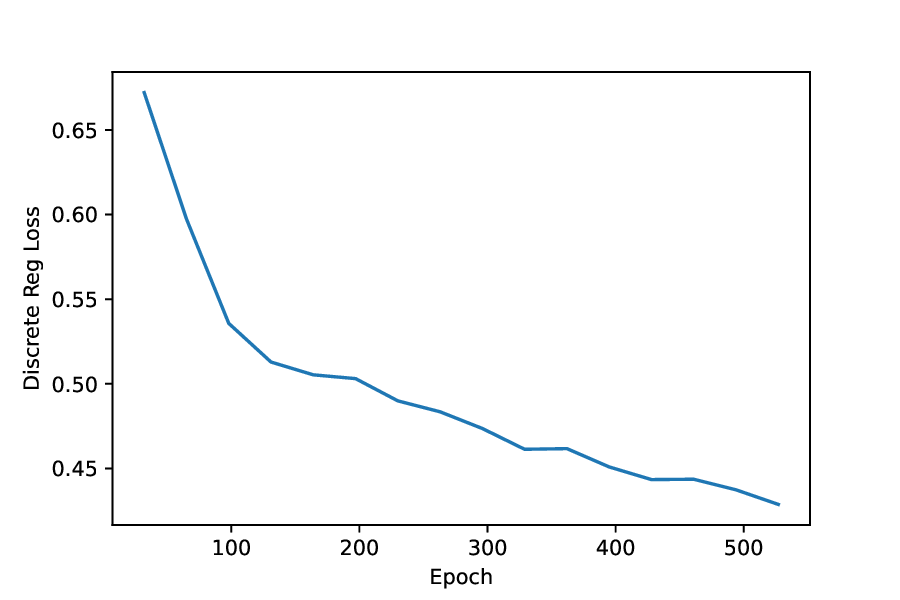}} 
\caption{Discrete Regularization and loss plot while finetuning FIGARO on EMOPIA data}
\end{figure}

\subsection{User Study Evaluation of ScripTONES}

\begin{figure}[htbp]
\centering
\subfigure[Box-plot of user ratings for E-CWT \& MVAE models\label{fig:box_plot}]{\includegraphics[width=0.35\columnwidth]{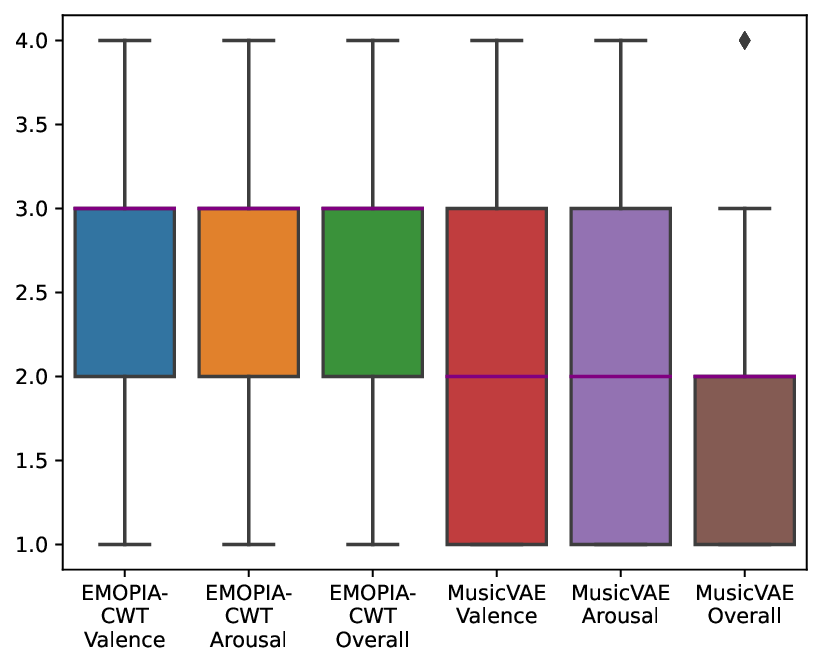}} 
\hfill
\subfigure[Music Knowledge rating of survey subjects\label{fig:music_knowledge}]{\includegraphics[width=0.4\columnwidth]{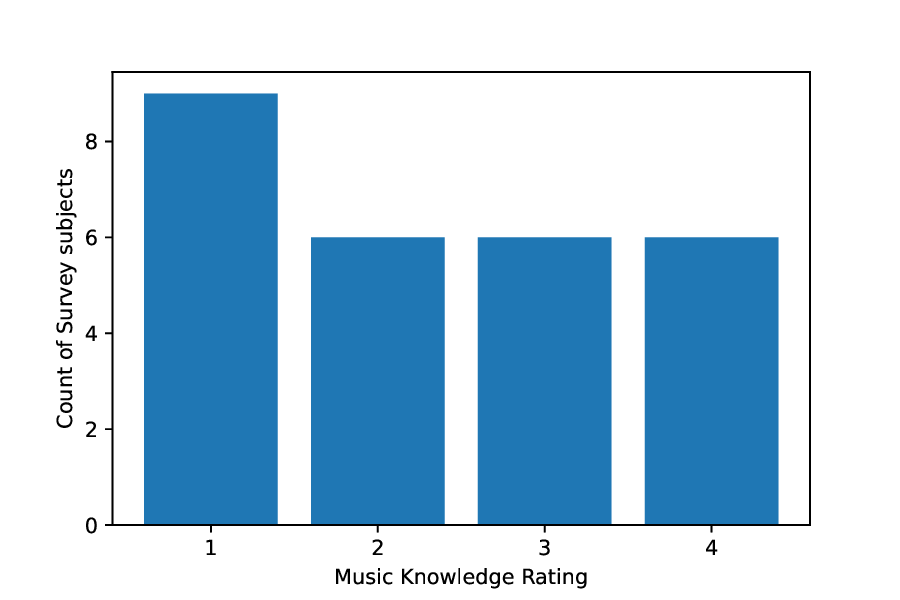}}

\caption{User preferences of music generation model and their music knowledge}
\end{figure}

A group of 31 users (with varying music knowledge as seen in Fig. \ref{fig:music_knowledge}) were surveyed to compare the quality of music generated by the different models for 3 movie scenes. For each scene, two different music pieces were played -- one generated by MusicVAE and the other generated by EMOPIA-CWT as the music generators in the ScripTONES pipeline. For each piece of music, on a scale of 1-4, we rate the valence/positivity, the arousal/excitement and the overall mood fit of the music with respect to the mood of the scene.

\begin{table}[]
\parbox{.45\linewidth}{
\centering
\caption{Average ratings of match between generated music and film scene}
\label{tab:user_study}
\begin{tabular}{c|c|c}
\hline
Attribute Rated & E-CWT & MVAE \\
\hline
Valence & \textbf{2.62} & 1.96 \\
Arousal & \textbf{2.44} & 1.92 \\
Overall Mood Fit & \textbf{2.48} & 1.86 \\
\hline
\end{tabular}
}
\hfill
\parbox{.45\linewidth}{
\centering
\caption{User evaluated scene-wise overall mood fit ratings on a scale of 1-4}
\label{tab:scene_wise}
\begin{tabular}{c|c|c}
\hline
Scene Number & E-CWT & MVAE \\
\hline
Scene 1 & \textbf{2.52} & 1.58 \\
Scene 2 & 1.87 & \textbf{2.17} \\
Scene 3 & \textbf{3.06} & 1.84 \\
\hline
\end{tabular}
}
\end{table}

Table \ref{tab:user_study} presents the average valence, arousal and overall mood fit ratings (on a 4-point Likert scale), across all users and all scenes in the study. A detailed box-plot of the ratings are shown in Fig. \ref{fig:box_plot}. The EMOPIA-CWT model receives a better rating across all attributes. This can be attributed to the compound word architecture of EMOPIA-CWT, which can generate more coherent and nuanced music with polyphony.

We analyse the scene-wise consensus (details in Appendix \ref{sec:app_scene}) among users by averaging the overall mood fit ratings for each scene across all users in Table \ref{tab:scene_wise}. We find that user's prefer EMOPIA-CWT in scenes that require nuanced and sophisticated music with polyphony whereas MusicVAE is preferred in negative and action scenes.

\subsection{Discrete Regularization} \label{sec:disc_reg}
To improve sentiment-conditioning further in VAEs we propose 
a Discrete Regularization loss as per Eqn. \ref{eq:disc_reg_main}. An illustration can be seen in Fig. \ref{fig:discrete_reg} and the regularization process is described in more detail in Appendix \ref{app:disc_reg}. We use this loss to finetune FIGARO's \cite{von2022figaro} pretrained VAE model, on the EMOPIA dataset \cite{hung2021emopia}. This helps FIGARO's VAE model learn a better latent space representation for sentiment as seen in Fig. \ref{fig:disc_reg_plot} and detailed further in Appendix \ref{app:finetune}.

\begin{equation} \label{eq:disc_reg_main}
    L_{reg_{disc}} = BCE(v_{pred}, v_{gt}) + BCE(a_{pred}, a_{gt})
\end{equation}

\goodbreak

\section{Conclusion and Future Work}
In this paper, we present ScripTONES, which introduces new directions for research on sentiment-conditioned music generation for movie scripts. We obtain the sentiment encoding of movie scripts using the NRC VAD lexicon for English words and use it to conditionally generate music using both VAE and Transformer architectures, which we qualitatively evaluate by surveying users. We present the ideas of attribute vector arithmetic and latent regularisation for sentiment modifications in the latent space of the VAE. In future work, variations of encoder-decoder architectures (GANs, GRU-VAEs) can be explored to improve the music generation capabilities of the model.

\bibliographystyle{plain}
\bibliography{refs.bib}

{
\small
}


\pagebreak

\appendix

\section{Improving Sentiment-Conditioning of Music with VAEs} \label{app:improving}

We propose a few methods to improve the affective characteristics of the music generated by a variational autoencoder (VAE) to match the sentiment of the script more closely. 
These methods are enabled by three properties of VAEs, which we describe below.
\smallskip

\subsection{Attribute Vector Arithmetic}
As described in Section \ref{sec:attributevec}, we can extract the characteristics for each type of sentiment in the embedding space and use it to tune the music generated. While we demonstrate the use of attribute vector arithmetic with the four quadrants of the valence-arousal space as our distinct sentiments, this can also be used for other \emph{distinct sentiment models} like positive-negative (2 categories), Plutchik's wheel of emotions 8 categories) \cite{plutchik1991emotions} and Ekman's model (6 categories) \cite{ekman1992argument}. 
However, we cannot extract attribute vectors for all sentiments in a continuous sentiment model.

\subsection{Latent Space Regularization}
To handle continuous sentiment models, we consider the work done by Pati et al. \cite{pati2021attribute}, which extended the ideas of latent space regularization \cite{engel2017latent} and disentanglement \cite{higgins2017beta} for music. The regularization loss is capable of constraining a particular dimension of the latent space to reflect monotonicity in a particular attribute of the data.
Regularizing one dimension of the latent space each for the attributes of valence and arousal will enable direct control of valence and arousal by simply increasing or decreasing the corresponding dimension. This is much like the faders described in \cite{tan2020musicfadernets} but for high-level sentiment attributes.

\smallskip

We propose regularizing the latent space in the following two ways. 

\subsubsection{\textbf{Continuous Regularization}}
The first, which we call \emph{Continuous Regularization} is closely related to \cite{pati2021attribute} where we define regularization losses for valence and arousal as per the equations
\begin{equation}
    L_{reg_{v}} = MSE(tanh(D_{z^1}) - sgn(D_{v}))
\end{equation}
\begin{equation}
    L_{reg_{a}} = MSE(tanh(D_{z^2}) - sgn(D_{a}))
\end{equation}
and $L_{reg_{cont}} =  L_{reg_{v}} + L_{reg_{a}}$, where $D_{z^1}$ and $D_{z^2}$ are the sum of differences in $z^1$ ($1^{st}$ dimension for valence) and $z^2$ ($2^{nd}$ dimension for arousal) respectively, computed pairwise over all pairs of data points in the mini-batch. $D_{v}$ and $D_{a}$ are the \emph{attribute distances} i.e. the sum of differences in the valence and arousal values of the pairs of data points. $sgn(.)$ refers to the sign (positive or negative).

\smallskip

Due to the unavailability of a dataset of music annotated with continuous valence and arousal values, computation of the \emph{attribute distances} is not directly possible. We suggest the use of the valence-arousal regressors introduced in MusAV \cite{bogdanov2022musav} to predict the valence and arousal values of each piece of music, thereby enabling the computation of $D_{a}$ and $D_{v}$. However, the valence-arousal predictions from the regressors can only be considered noisy labels (true labels in \cite{bogdanov2022musav} are relative valence-arousal and not absolute values) and might lead to inaccurate regularization. 

\smallskip

\subsubsection{\textbf{Discrete Regularization}} \label{app:disc_reg}
To use true labels instead of noisy predictions as labels for regularization, we propose a second method for regularization called \emph{Discrete Regularization}, inspired by \cite{nath2021levasa}. Here, we use the discrete VA quadrant annotations in the EMOPIA dataset \cite{hung2021emopia} as an approximation for the continuous VA values. 

\smallskip 

Here again, we regularize $z^1$ for valence and $z^2$ for arousal. This is done by passing the $z^1$ and $z^2$ through a simple single hidden layer neural network with a single output neuron. Sigmoid activation is applied to the output neuron to predict the probability of whether the corresponding attribute (valence or arousal) is high (1) or low (0). The regularization loss is formulated as 
\begin{equation} \label{eq:disc_reg}
    L_{reg_{disc}} = BCE(v_{pred}, v_{gt}) + BCE(a_{pred}, a_{gt})
\end{equation}
where $BCE(.)$ refers to Binary Cross Entropy Loss, $v_{pred}$ and $a_{pred}$ are the sigmoid probabilies for valence and arousal values. $v_{gt}$ and $a_{gt}$ are ground truth values for valence and arousal based on the quadrant annotations. For instance, for a music piece with annotation as quadrant 4, the valence is high and the arousal is low. This implies $v_{gt} = 1 $ and $a_{gt} = 0$. An illustration of Discrete Regularization can be seen in Fig. \ref{fig:discrete_reg}.

\smallskip

Once either Discrete or Continuous regularization has been performed during training, the valence and arousal attributes can be manipulated by simply changing the values of the dimensions $z^1$ and $z^2$ for any piece of music. This is done either as per $VA_i$ extracted from the script or a slider, by performing a simple addition $z^1 = z^1 + V$ and $z^2 = z^2 + A$ before decoding the music.

\subsection{Latent Space Interpolation}

\begin{figure}[h]
\centering
\includegraphics[width=0.4\columnwidth]{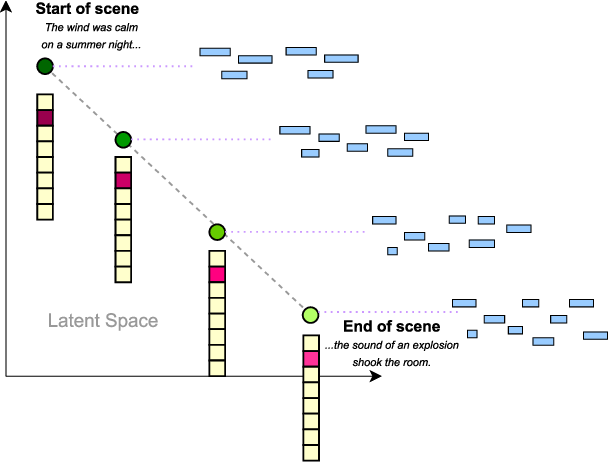}
\caption{Latent Space Interpolation}
\label{fig:interpolation}
\end{figure}

Since attributes of the data are encoded within the latent space by the VAE, interpolation in the latent space of the VAE would allow us to smoothly change the attributes in a piece of music. This would be particularly useful when the sentiment changes over the course of a scene. For instance, the start of the scene could have low arousal while the end of the scene has high arousal. The changes in sentiment can be captured by extracting $VA$ values at a sentence level and manipulating $z$ for the start and end of a scene using $VA_{start}$ and $VA_{end}$ which refer to the $VA$ values for the first and last sentence of a scene respectively. The pieces of music for the other parts of the scene can be filled using interpolation between $z_{start}$ and $z_{end}$, as can be seen in Fig. \ref{fig:interpolation}.

\section{Scene-Wise Analysis of User Survey} \label{sec:app_scene}
We analyse the scene-wise consensus among users by averaging the overall mood fit ratings for each scene across all users in Table \ref{tab:scene_wise}. Scene 1 --- a dramatic, high-impact and large-scale sequence --- requires nuanced and sophisticated music to enhance the experience, which EMOPIA-CWT is able to do better with its polyphony. In Scene 2 --- which is a negative, action-packed fight scene --- MusicVAE does better, as EMOPIA-CWT is known to struggle with music with negative sentiment. Additionally, the action in the scene is paralleled better in MusicVAE due to attribute vector arithmetic, where the addition of the high arousal vector ($z_{ah}$) induces fast-paced notes in music. EMOPIA-CWT again does better in Scene 3 as it is an emotional scene with almost no action sequence.

\section{Additional Experiments} \label{sec:add_exp}
\subsection{Finetuning VAE with Discrete Regularization} \label{app:finetune}
FIGARO \cite{von2022figaro} is a recent work that uses VAEs to extract high-level "descriptions" at a bar-level, and then uses the sequence of descriptions to generate coherent music with a Seq2Seq Transformer model.
We finetune FIGARO's \cite{von2022figaro} pretrained VAE model, on the EMOPIA dataset \cite{hung2021emopia}. This is done by adding a discrete regularization component to the VAE training loss as per Eqn \ref{eq:disc_reg}, to improve sentiment conditioning (described in \ref{sec:disc_reg}). This helps FIGARO's VAE model learn a better latent space representation for sentiment, in addition to the reconstruction properties carried over from the pretraining.

In Fig. \ref{fig:disc_reg_plot}, the discrete regularization loss continues to decrease during finetuning which indicates that the model is able to predict the Valence and Arousal values of the piece of music better after finetuning. However, the finetuning process is very slow and takes many epochs to converge as the gradient from the discrete regularization loss component is much smaller relative to the reconstruction and KL divergence loss components of the VAE.

\subsection{Polyphonic Recurrent VAE} \label{app:polyvae}
Experiments were conducted to train a simple Recurrent VAE baseline for polyphonic music with a similar architecture to \cite{bowman2015generating}. We use a bidirectional LSTM as the encoder and a vanilla LSTM as the decoder, with one layer each. During training, the input music is fed into the bi-LSTM encoder. The hidden vector of the last state of the bi-LSTM is mapped to the latent dimensions through a fully connected layer, which is then fed into the LSTM decoder to output music. We additionally concatenate the latent vector $z$ with the input embedding at every time step of the decoder. 


The polyphonic recurrent VAE however suffers from posterior collapse, where the LSTM decoder simply ignores the latent vector $z$ and learns to output a repeating sequence of notes. This is especially the case when teacher forcing rate \footnote{percentage of times ground truth is fed back as input to decoder instead of previous decoder output}
is set to 0 in the decoder. When the teacher forcing rate is set to 1, the problem of repeating notes is solved, but the reconstruction for all pieces of music remains the same due to posterior collapse.

We detail the results of all our experiments, as well as the hyperparameters chosen in the Supplementary Material.

\section{Related Work} \label{app:related_work}

\subsection{Sentiment Analysis}

One-dimensional sentiment polarity analysis, commonly referred to as sentiment analysis, refers to identifying whether textual data conveys a positive or negative sentiment. Many benchmark datasets for sentence-level and phrase-level binary classification  \cite{socher2013recursive, maas2011learning, yelp, ni2019justifying} have been developed. Pre-trained Transformer-based models \cite{raffel2020exploring, devlin2018bert, yang2019xlnet, liu2019roberta} have been fine-tuned on the benchmark datasets to obtain state-of-the-art binary sentiment classification.

To capture more complex sentiments, sentiment analysis techniques based on discrete sentiment categories have been explored. Few public datasets exist for multi-emotion classification. EmoLex \cite{mohammad2013crowdsourcing} maps English language words to 8 discrete emotion categories. ISEAR \cite{scherer1994evidence} and GoEmotions \cite{demszky2020goemotions} are multi-emotion sentence classification datasets with 28 and 7 emotion classes respectively.

The NRC VAD \cite{vad-acl2018} lexicon maps 20,000 English words to a continuous space of real-valued valence, dominance and arousal scores. For sentence-level continuous emotion regression, the EmoBank \cite{buechel-hahn-2017-emobank} dataset contains 10,000 sentences labelled with valence, arousal and dominance values. The drawback of this dataset is that there is an imbalance wherein most sentences are labelled with VA values in the range of 2 to 3.

\subsection{Music Generation}
Music generation has been approached in two major ways. The first is to generate music as raw audio (a \verb|.wav|, for instance) \cite{dhariwal2020jukebox, agostinelli2023musiclm}. The second is to generate music in the symbolic space either as a probability over all possible notes for each time step \cite{roberts2018hierarchical, huang2018musictransformer} or as a sequence of tokens in one of many tokenized representations of music \cite{midilike2018, remi2020, octuple2021, hsiao2021compound}, which can then be converted to a MIDI (\verb|.mid|) file. We discuss approaches for symbolic music generation due to the complexities associated with raw audio domain music generation.

Variational autoencoders for generating music were popularised by MusicVAE \cite{roberts2018hierarchical}, a hierarchical variational autoencoder with recurrent neural networks (RNNs) as encoder and decoder. 

To decompress the latent representation while avoiding posterior collapse, a hierarchical RNN decoder was used. However, the output produced by the MusicVAE architecture is limited to \textit{monophonic music} and is unable to produce \textit{polyphonic music} \footnote{In monophonic music, only one note can be played at any time. In polyphonic music, multiple notes can be played together simultaneously (like chords)}.

The authors also proposed using attribute vector arithmetic in the latent space but limited it to simpler attributes like note density. More recent VAE approaches for music like \cite{wang2020pianotree, jiang2020transformervae, von2022figaro} have handled polyphonic music in a VAE. In addition to VAEs, transformers \cite{vaswani2017attention} have also been used widely to generate music \cite{huang2018musictransformer, hsiao2021compound}.


Various approaches have recently been proposed specifically for sentiment-conditioned music generation. Lucas et al. \cite{ferreira2021learning} propose a combination of mLSTM and a logistic regression model. The MIDI files are represented as a series of events, relevant to features such as timbre, harmony, tempo, etc. The sentiment is interpreted from these features to create the partially annotated VG-MIDI dataset (annotated with a valence-arousal pair). The generative mLSTM is trained on the unlabeled dataset and the logistic regressor uses the hidden state to encode MIDI phrases and predict statements. However, it fails to accurately generate music for negative sentiments and cannot handle inputs more detailed than a given set of labels/emotions.


Hung et al. \cite{hung2021emopia} introduced the EMOPIA dataset, a collection of labelled pop-piano midi files based on the valence-arousal model of sentiment. The authors provided baselines for music classification and sentiment-conditioned music generation. The compound word transformer (CWT) \cite{hsiao2021compound} was the baseline proposed as a sentiment-conditioned music generator. This however is limited to generating music conditioned only by a given quadrant of the VA space. 
.


Other recent approaches to this problem include Transformer-GANs \cite{neves2022transformergan} and Music FaderNets \cite{tan2020musicfadernets}. The first proposes using an additional adversarial loss component to the usual negative log-likelihood (NLL) loss utilized to train transformers for music generation, as a way to prevent the degradation of quality in longer sequences of music. However, the conditioning control is again only limited to the quadrant number and does not allow for sentiment modifications to a given piece of music. Music FaderNets aimed at bypassing the question of formalizing high-level notions like "sentiment" by trying to force the model to infer them from a dataset of low-level features using semi-supervised clustering. Currently, however, FaderNets are trained only to be conditioned on arousal.

\section{Data and Dataset Preprocessing} \label{app:data}
For movie scripts, we use the IMSDb \cite{imsdb} and MovieNet \cite{huang2020movienet} databases to source movie scripts. For sentiment analysis of text, we use the NRC VAD \cite{vad-acl2018} lexicon, a dataset of valence, dominance and arousal human ratings. For music generation tasks, we use the AILabs Pop1k7 dataset \cite{hsiao2021compound} and the EMOPIA dataset \cite{hung2021emopia} which contains MIDI piano pieces categorized into one of the 4 quadrants in the valence-arousal space. 

We used the MidiTok \cite{miditok2021} library to tokenize the AILabs1k7 and EMOPIA datasets into REMI \cite{remi2020} representations, and chunked them into single-bar representations.

\end{document}